\documentclass[twocolumn,english,aps,prc,amsmath,amssymb,showpacs]{revtex4}

\usepackage[T1]{fontenc}
\usepackage[latin9]{inputenc}
\usepackage{babel}

\usepackage{graphicx}

\begin{document}

\title{Topics in nuclear structure}

\author{A. Escuderos}
\author{L. Zamick}

\affiliation{Department of Physics and Astronomy, Rutgers University, Piscataway,
New Jersey, 08854 USA}

\begin{abstract}
We consider work performed over the last decade on single-$j$-shell studies. We 
will discuss four topics.
\end{abstract}

\maketitle

\section{A study of the E$(J_{\text{max}})$ interaction}

This is an extension of work done with Apolodor Raduta and Elvira Moya de~Guerra
when they visited Rutgers on a NATO grant in 2001~\cite{mrzs03}. In contrast to work on
$J=0$ pairing, we here study, in a single $j$ shell, an interaction $E(J_{\text{max}})$
for which all two-body matrix elements vanish except for $J=J_{\text{max}}=2j$. 
In this work we will focus on the model itself. Comparisons with realistic interactions 
are given in Ref.~\cite{ze13}

For a two-particle (or two-hole) system in a single $j$ shell, the even-$J$ states have 
isospin $T=1$ (triplets) and the odd-$J$ states have isospin $T=0$ (singlets). Thus, 
an interaction acting in only $J_{\text{odd}}$ states, including $J=J_{\text{max}}$, 
cannot occur for two neutrons or two protons in a single $j$ shell---only in the 
neutron--proton system. This leads to some simplifications.

We will consider a system of two protons and two neutrons (or holes), e.g. $^{96}$Cd
in the $g_{9/2}$ shell. The basis states are $| [pp(J_p) nn(J_n)]^{I} \rangle$, where 
$I$ is the total angular momentum. To satisfy the Pauli principle, $J_{p}$ and $J_{n}$
must be even.

For $I=0$, the secular Hamiltonian is separable and $J_n=J_p$:
\begin{equation}
H_{J_{p},J_{p'}}=V(J_{\text{odd}})f(J_{p})f(J_{p'})\,,
\end{equation}
where $f(J_{p})$ is twice the unitary 9-$j$ symbol ($U9$-$j$): 
\begin{equation}
\begin{aligned}
f(J_{p}) & = 2\langle(jj)^{J_{p}}(jj)^{J_{p}}|(jj)^{J_{\text{odd}}}
(jj)^{J_{\text{odd}}}\rangle^{I=0}= \\
 & = 2(2J_{p}+1)(2J_{\text{odd}}+1)
 \begin{Bmatrix}
 j & j & J_{p}\\
j & j & J_{p}\\
J_{\text{odd}} & J_{\text{odd}} & 0
\end{Bmatrix}
\end{aligned}
\end{equation}

If we write the wave function as $\sum X_{J_{p}J_{p}} |[pp(J_{p}) nn(J_{p})]^{I=0}\rangle$, 
then it was shown in Ref.~\cite{mrzs03} that $X_{J_{p}J_{p}}$ is proportional 
to $f(J_{p})$. The other $I=0^{+}$ eigenstates are degenerate and, if 
$V(J_{\text{odd}})$ is negative, they are at higher energies. In other words, 
what we have shown in Ref.~\cite{mrzs03} is that the wave-function components 
$X_{J_{p}J_{p}}$ of the lowest $I=0^{+}$ state are equal within a normalization 
to the $U9$-$j$ coefficients: $\sqrt{2} \langle (jj)^{J_{p}} (jj)^{J_{n}} | 
(jj)^{J_{\text{max}}} (jj)^{J_{\text{max}}} \rangle^{I=0}$. 

The eigenvalue is given by 
\begin{equation}
E(I=0^{+})=V(J_{\text{odd}})|\sum f(J_{p})X_{J_{p}J_{p}}|^{2}
\end{equation}
Note that our very simple interactions are charge independent. This
means that the lowest (non-degenerate) $I=0^{+}$ state has good isospin,
presumably $T=0$. It is amusing that we can assign the isospin quantum
number to a wave function with $U9$-$j$ coefficients.

Similarly, we can show that there is also a very simple expression for the 
$I=1^{+}$ lowest eigenfunction: $2 \langle (jj)^{J_{p}} (jj)^{J_{n}} | 
(jj)^{J_{\text{max}}} (jj)^{J_{\text{max}}-1} \rangle^{I=1}$.

However, for states of angular momentum 2 or higher, the secular Hamiltonian is 
no longer separable. The eigenvalue equation is
\begin{multline}\label{eq:i2ev}
4\sum_{J_{x}}\langle(jj)^{J_{p}}(jj)^{J_{n}}|(jj)^{9}(jj)^{J_{x}}\rangle^{I} \times \\
\times \sum_{J_{p'}J_{n'}}\langle(jj)^{J_{p'}}(jj)^{J_{n'}}|
 (jj)^{9}(jj)^{J_{x}}\rangle^{I} X_{J_{p'}J_{n'}} = \\
= \lambda X_{J_{p} J_{n}} \,,
\end{multline}
where $\lambda$ is the eigenvalue and $X_{J_{p} J_{n}}$ stands for the
eigenfunction components. For $I=2^{+}$ there are two terms corresponding 
to $J_{x}=7$ and 9; for $I=3^{+}$ the values are $J_{x}=6$ and 8, etc.

Despite the complexity of the above equation, there are some surprising
results. The eigenfunction components of the lowest $2^{+}$ state
are numerically extraordinarily close to the single $U9$-$j$ symbols
$\sqrt{2}\langle(jj)^{J_{p}}(jj)^{J_{n}}|(jj)^{9}(jj)^{9}\rangle^{I=2}$.
Furthermore, the next $2^{+}$ state has also components exceedingly
close to $2\langle(jj)^{J_{p}}(jj)^{J_{n}}|(jj)^{9}(jj)^{7}\rangle^{I=2}$.
This is by no means obvious because, as mentioned above, the interaction
involves a sum of two separable terms corresponding to $J_{x}=7$
and 9. 

We can explain this result by performing a calculation of the overlap of
the two $U9$-$j$'s of the last paragraph. We restrict the sum to even 
$J_{p}$ and even $J_{n}$. We first note schematically 
\begin{multline}\label{eq:sch}
4\sum_{\text{even }J_{p},J_{n}} = \sum(1+(-1)^{J_{p}})(1+(-1)^{J_{n}})= \\
 = \sum+\sum(-1)^{J_{p}}+\sum(-1)^{J_{n}}+\sum(-1)^{J_{p}+J_{n}}
\end{multline}

Using orthogonality relations for 9-$j$ symbols, we can see that the first 
term vanishes. In the last term, one of the $U9$-$j$'s has two rows that 
are the same, which means that the only non-vanishing terms in the sum have 
($J_{p}+J_{n}$) even. Thus, the last term is the same as the first---zero. The
two middle terms are the same, so we get the overlap of the two $U9$-$j$'s to be
\begin{subequations}
\begin{align}
\mathop{\sum_{\text{even }}}_{J_{p},J_{n}} = & \frac{1}{2}
\mathop{\sum_{\text{even }}}_{J_{p},J_{n}} (-1)^{J_{p}}\langle(jj)^{J_{p}}
(jj)^{J_{n}}|(jj)^{9}|(jj)^{9}\rangle^{I=2}\times \\
 & \times\langle(jj)^{J_{p}}(jj)^{J_{n}}|(jj)^{9}(jj)^{7}\rangle^{I=2}= \\
 = & -\frac{1}{2}\langle(jj)^{9}(jj)^{9}|(jj)^{9}(jj)^{7}\rangle^{I=2}\,.
 \label{eq:i2}
\end{align}
\end{subequations}
We obtain the above by using again orthogonality relations for $9j$-symbols. 

Using similar arguments, one can show that the normalization for the 
$|[pp(9)nn(9)]^{I=2}\rangle$ state is such that its normalization factor is
\begin{equation}\label{eq:n2-99}
\begin{aligned}
N(9)^{-2} & = \frac{1}{2}-\frac{1}{2}\langle(jj)^{9}(jj)^{9}|
(jj)^{9}(jj)^{9}\rangle^{I=2}=\\
 & = \frac{1}{2}-\frac{1}{2}0.00001209813=0.499993950935
\end{aligned}
\end{equation}

For the $|[pp(9)nn(7)]^{I=2}\rangle$ state, we obtain 
\begin{equation}\label{eq:n2-97}
\begin{aligned}
N(7)^{-2} & = \frac{1}{4}-\frac{1}{2}\langle(jj)^{9}(jj)^{7}|
(jj)^{9}(jj)^{7}\rangle^{I=2}= \\
 & = \frac{1}{4}+\frac{1}{2}0.00075253477=0.250376267385
\end{aligned}
\end{equation}

To get this latter result, we use the following relationship 
\begin{equation}
\sum_{J_p,J_n}(-1)^{(J_{p}+J_{n})}\left|\langle(jj)^{9}(jj)^{7}|
(jj)^{J_{p}}(jj)^{J_{n}}\rangle^{I=2}\right|^{2}=0
\end{equation}

From Eqs.~\eqref{eq:n2-99} and \eqref{eq:n2-97}, we find that the 
normalizations are 1.414222 and 1.998497, only slightly different 
from $\sqrt{2}$ and 2 respectively. Therefore, we obtain that the 
term in Eq.~\eqref{eq:i2} is exceedingly small for the $g_{9/2}$ 
shell, namely 0.00009113 and, if we include the exact normalization factors, 
we get 0.00025756.

We can see in Table~\ref{tab:j2} that the results for matrix diagonalization
for both $I=2^{+}$ states yield wave function components which are
very close to the normalized $U9$-$j$ coefficients. In fact, they
are so close that one could wonder if they are exactly the same. But
they are not. As seen in Eq.~\eqref{eq:i2}, the two $U9$-$j$ sets
corresponding to $J_x=9$ and $J_x=7$ are very nearly orthogonal,
but not quite.

\begin{table}[h!t]
\caption{Comparison for the first two $I=2^{+}$ states of
the matrix diagonalization with the E(9) interaction and with normalized
$U9$-$j$ components. We give the energy in MeV in the second row.}
\centering
\begin{tabular}{|c|c|c|c|c|}
\hline
$[J_{p}\,,J_{n}]$  & E(9)  & $U9$-$j$  & E(9)  & $U9$-$j$ \cr
 & 1.069  &  & 3.0558  & \cr
\hline 
$[0\,,2]$  & 0.5334  & 0.5338  & 0.1349  & 0.1351 \cr
$[2\,,2]$  & $-0.4707$  & $-0.4708$  & 0.5569  & 0.5567 \cr
$[2\,,4]$  & 0.3035  & 0.3035  & 0.3188  & 0.3189 \cr
$[4\,,4]$  & $-0.1388$  & $-0.1390$  & 0.6300  & 0.6299 \cr
$[4\,,6]$  & 0.0531  & 0.0531  & 0.1320  & 0.1320 \cr
$[6\,,6]$  & $-0.0137$  & $-0.0138$  & 0.1350  & 0.1350 \cr
$[6\,,8]$  & 0.0025  & 0.0025  & 0.0114  & 0.0114 \cr
$[8\,,8]$  & $-0.0003$  & $-0.0003$  & 0.0052  & 0.0052 \cr
\hline
\end{tabular}
\label{tab:j2}
\end{table}

It turns out that all the other lowest even-$I$ states have eigenfunctions
close although not exactly equal to $\sqrt{2}\langle(jj)^{J_{p}}(jj)^{J_{n}}|
(jj)^{9}(jj)^{9}\rangle^{I}$. In Table~\ref{tab:9j} we compare, as an 
example, the wave function of the $I=8^{+}$ state. In the second column, 
we give the single $U9$-$j$ symbols (normalized) and in the third column 
we give results of diagonalizing the E(9) interaction. Since the coefficient 
$[J_{p},J_{n}]$ is the same as $[J_{n},J_{p}]$, we list only one of them. 
The overlap of the two wave funtions is 0.9998.

\begin{table}[h!t]
\caption{Comparing the wave functions of a single $U9$-$j$
symbol with $J_{x}=9$ with a full diagonalization of E(9) for the
lowest $I=8^{+}$ state in $^{96}$Cd.}
\centering
\begin{tabular}{|c|c|c|}
\hline
$[J_{p}\,,J_{n}]$  & $U9$-$j$  & E(9) \cr
\hline 
$[0\,,8]$  & 0.0630  & 0.0644 \cr
$[2\,,6]$  & 0.4299  & 0.4271 \cr
$[2\,,8]$  & $-0.0522$  & $-0.0513$ \cr
$[4\,,4]$  & 0.7444  & 0.7456 \cr
$[4\,,6]$  & $-0.1803$  & $-0.1729$ \cr
$[4\,,8]$  & 0.0256  & 0.0280 \cr
$[6\,,6]$  & 0.0521  & 0.0657 \cr
$[6\,,8]$  & $-0.0076$  & $-0.0012$ \cr
$[8\,,8]$  & 0.0011  & 0.0047 \cr
\hline
\end{tabular}
\label{tab:9j}
\end{table}

We can plot the coupling of Eq.~\eqref{eq:i2} for various shells, as 
we can see in Table~\ref{tab:u9j-sh}. We see that the coupling $U9$-$j$ 
decreases at least exponentially as we go up in shell. Experts say 
that this type of 9-$j$ lies in the non-classical region.

\begin{table}[h!t]
\caption{Value of coupling $U9$-$j$ symbols for various shells.}
\centering
\begin{tabular}{|c|c|c|}
\hline
$j$ & $U9$-$j$ & overlap of Eq.~\eqref{eq:i2} \cr
\hline
$p_{3/2}$ & $-0.1800$ & 0.2546 \cr
$d_{5/2}$ & $-0.021328$ & 0.03016 \cr
$f_{7/2}$ & $-0.002074$ & 0.002933 \cr
$g_{9/2}$ & $-0.0001822$ & 0.0002577 \cr
$h_{11/2}$ & $-0.00001502$ & 0.00002174 \cr
$i_{13/2}$ & $-0.000001185$ & 0.000001676 \cr
\hline
\end{tabular}
\label{tab:u9j-sh}
\end{table}

\subsection{Degeneracies}

With the E$(J_{\text{max}})$ interaction for $g_{9/2}$ [that is, E(9)], 
we get several degenerate states with an
absolute energy zero. In some detail, for $I=0^{+}$ there are five
states, three with isospin $T=0$ and two with $T=2$. There is one
non-degenerate state at an energy $2V(9)$ ($V(9)$ is negative).
The other four $I=0^{+}$ states have zero energy. For $I=1^{+}$
all states have isospin $T=1$. There is a single non-degenerate state
at $V(9)$, the other three have zero energy. For $I=2^{+}$ there
are twelve states---six have $T=0$, four have $T=1$, and two have
$T=2$. There are two non-degenerate $T=0$ states with approximate
energies $2V(9)$ and $V(9)$ respectively, and one non-degenerate
$T=1$ state with energy $V(9)$. The other nine states have zero
energy. To understand this, take a wave function 
\begin{equation*}
|\Psi^{\alpha}\rangle = \sum_{J_p, J_n}{C^{\alpha}(J_{p},J_{n})|
[pp(J_{p}) nn(J_{n})]^{I}\rangle}
\end{equation*}
and the corresponding energies $E^{\alpha}=\langle\Psi^{\alpha}
|H|\Psi^{\alpha}\rangle$. Consider the sum $\sum_{\alpha}{E^{\alpha}}$. 
We have 
\begin{equation}
\sum_{\alpha}{C^{\alpha}(J_{p},J_{n})C^{\alpha}(J_{p'},J_{n'})}=
\delta_{J_{p},J_{p'}}\delta_{J_{n},J_{n'}}
\end{equation}
Thus 
\begin{equation}\label{eq:sume}
\begin{aligned}
\sum_{\alpha}{E^{\alpha}} & = \sum_{J_{p}J_{n}}{\langle[pp(J_{p}) nn(J_{n})]^{I}|
H|[pp(J_{p}) nn(J_{n})]^{I}\rangle} = \\
 & = 4V(9)\mathop{\sum_{J_{p}J_{n}}}_{\text{even}}{\sum_{J_{x}}{\left|
 \langle(jj)^{J_{p}}(jj)^{J_{n}}|(jj)^{9}(jj)^{J_{x}}\rangle^{I}
 \right|^{2}}} 
\end{aligned}
\end{equation}

This expression does not depend on the detailed wave functions. Referring
to Eqs.~\eqref{eq:n2-99} and \eqref{eq:n2-97} and neglecting the very small 
correction terms, we see that $N^{-2}$ is equal to $1/2$ for $J_{x}=9$ and 
to $1/4$ for all other $J_{x}$. Basically then Eq.~\eqref{eq:sume} becomes 
$4 V(9) \sum N(J_{x})^{-2}$. Hence we obtain $\sum_{\alpha}{E^{\alpha}}=2V(9)$ for 
$I=0$, $V(9)$ for $I=1$, and $4V(9)$ for $I=2$. But we can alternately
show, using the explicit wave functions, that for $I=0$ the energy
of the lowest state is $2V(9)$. Hence, all the other states must
have zero energy. A similar story for $I=1$. The $I=2$ state is
a bit more complicated because of the coupling between two states,
however small it is. Still one can work it through and see that the
$4V(9)$ energy is exhausted by the two $T=0$ and the one $T=1$
non-degenerate states.

For $I=0$ we have two $T=0$ and two $T=2$ states, all degenerate. One can remove
the degeneracies of $T=0$ and $T=2$ by adding to the Hamiltonian an interaction
$b\; t(i)\cdot t(j)$. This will not affect the wave functions of the non-degenerate
states but will shift the $T=2$ states away from the formerly degenerate
$T=0$ states.

The number of neutron-proton pairs with even $J$ for a system of two neutrons and
two protons is given by~\cite{z07b}
\begin{equation}
\sum_{J_{a}} | D^{I}(J_{a} J)| ^{2} (\delta_{T,0}
+ 4 \delta_{T,2}) ~,
\end{equation}
where $D^{I}(J_{P} J_{N})$ is the probability amplitude that in a state of total 
angular momentum $I$ , the protons couple to $J_{P}$ and the neutrons to $J_{N}$.

\section{How to handle degeneracies---pairing and $Q\cdot Q$}

We show how degeneracies, accidental or otherwise, can obscure some
interesting physics. But we further show how one can get around this problem. 

In a 2006 publication, Escuderos and Zamick~\cite{ez06} found some
interesting behaviour in the $g_{9/2}$ shell. Unlike the lower shells,
e.g. $f_{7/2}$, seniority is not a good quantum number in the $g_{9/2}$
shell. Despite this, it was found that in a matrix diagonalization
with four identical particles in the $g_{9/2}$ shell with total angular
momentum $I=4$ or 6, one unique state emerged no matter what interaction
was used. This problem was also addressed by others. 
Before the mixing, one has two states with seniority $v=4$ 
and one with $v=2$. The surprise was that, after the diagonalization, 
one gets a unique state that is always the same independently of the 
interaction used. This unique state has seniority $v=4$. The components 
of the wave function are given in the fourth column of Table~\ref{tab:res}
(labelled ``$T=2,$ $v=4$ unique"). The problem to be dealt with was not only 
why this state did not mix with the $v=2$ state, but also why it does 
not mix with the other $v=4$ state. But this will not concern us 
here. Rather we will use this as an example of how degeneracies can 
obscure interesting physics.

We first consider how the unique $T=2$ wave function looks like for a 
system of three protons and one neutron. This is shown in Table~\ref{tab:uni}.
No matter what interaction is used, this appears as a unique state.

\begin{table}[h!t]
\caption{A unique $J=4$, $v=4$ cfp for $j=9/2$.}
\centering
\begin{tabular}{|l|c|}
\hline
$J_{0}$ & $(j^{3} J_{0} j |\} I=4, v=4)$ \cr
\hline
3/2 & 0.1222 \cr
5/2 & 0.0548 \cr
7/2 & 0.6170 \cr
9/2 ($v=1$) & 0.0000 \cr
9/2 ($v=3$) & 0.0000 \cr
11/2 & $-0.4043$ \cr
13/2 & $-0.6148$ \cr
15/2 & $-0.1597$ \cr
17/2 & 0.1853 \cr
\hline
\end{tabular}
\label{tab:uni}
\end{table}

It was already commented on in a later paper by Zamick~\cite{z07} that cfp's
for identical particles are usually calculated using a pairing interaction.
With such an interaction, the two $v=4$ states are degenerate, i.e. they have 
the same energy. This means that any linear combination of the two states can 
emerge in a matrix diagonalization. Thus, the emergence of a unique state gets 
completely lost. The problem was also addressed in Refs.~\cite{zv08,q11,ih08}.

In this work we consider a less obvious example: a matrix diagonalization
of two proton holes and two neutron holes in the $g_{9/2}$ shell,
i.e. we consider $^{96}$Cd rather than $^{96}$Pd, the latter consisting
of four proton holes (whether we consider holes or particles does
not matter). We use a quadrupole--quadrupole interaction $Q\cdot Q$
for the matrix diagonalization. The two-body matrix elements in units
of MeV from $J=0$ to $J=9$ are: $-1.0000$, $-0.8788$, $-0.6516$,
$-0.3465$, $-0.0152$, 0.2879, 0.4849, 0.4849, 0.1818, and $-0.5454$.

We show some relevant results in Table~\ref{tab:res}. For $I=4$,
we get 14 eigenfunctions, but we list only two of them in the first
two columns. The reason we single these out is that they are degenerate---both
are at an excitation energy of 3.5284~MeV.

In the third wave function column, we have the unique state, one that
emerges, as we said above, with any interaction, however complicated,
e.g. CCGI~\cite{ccgi12}. But now we have to modify the phrase ``any
interaction''. We do not see this unique state when we use the $Q\cdot Q$
interaction---none of the 14 states looks like the one in column 3.
Learning from our experience with the pairing interaction, we suspect
that the problem lies with the two degenerate states at 3.5284~MeV.
We assumed that the two states were mixtures of one $T=0$ and one
$T=2$ state.

\begin{table*}[h!t]
\caption{Selected $I=4^{+}$ states in $^{96}$Cd with a $(g_{9/2})^{4}$
configuration. On the second row we give the energies in MeV.}
\centering
\begin{tabular}{|c|c|c|c|c|c|}
\hline
 & Mix $T=0,2$  & Mix $T=0,2$  & $T=2$, $v=4$ unique  & $T=0$ untangled  & other $T=2$, $v=4$ \cr
 & 3.5284  & 3.5284  & 6.5285  & 3.5284  & \cr
$[J_{p}\,,J_{n}]$  &  &  &  &  & \cr
\hline
$[0\,,4]$  & 0.0000  & 0.0000  & 0.0000  & 0.0000  & 0.0000 \cr
$[2\,,2]$  & $-0.3250$  & $-0.4170$  & $-0.4270$  & 0.3123  & $-0.0255$ \cr
$[2\,,4]$  & $-0.2364$  & $-0.2472$  & $-0.2542$  & 0.2289  & $-0.1986$ \cr
$[2\,,6]$  & 0.2168  & 0.3043  & 0.3107  & $-0.2076$  & $-0.1976$ \cr
$[4\,,4]$  & 0.0207  & 0.2390  & 0.2395  & $-0.0135$  & $-0.3313$ \cr
$[4\,,6]$  & $-0.1826$  & $-0.1364$  & $-0.1418$  & 0.1784  & 0.2245 \cr
$[4\,,8]$  & 0.0934  & 0.1540  & 0.1567  & $-0.0888$  & 0.3874 \cr
$[6\,,6]$  & $-0.1312$  & 0.1678  & 0.1638  & 0.1362  & 0.5645 \cr
$[6\,,8]$  & $-0.1343$  & 0.0357  & 0.0316  & 0.1353  & 0.0247 \cr
$[8\,,8]$  & $-0.7421$  & 0.5881  & 0.5625  & 0.7594  & $-0.1087$ \cr
\hline
\end{tabular}
\label{tab:res}
\end{table*}

We can remove the degeneracy without altering the wave functions of
the non-degenerate states by adding a $t(1)\cdot t(2)$ interaction
to the Hamiltonian. This will shift energies of states of different
isospin. What we actually did was equivalent to this. We added $-1.000$~MeV
to the two-body $T=0$ matrix elements. These had odd spin $J=1,3,5,7,9$.
What emerged is shown in columns 3 and 4. The degeneracy is removed.
We have a $T=2$ state in the third column shifted up by 3~MeV and
in the fourth column a $T=0$ state unshifted. The wave function components
are different from what they are in the first two columns. The $T=2$
state is the unique state we were talking about---one that emerges
with any interaction, e.g. CCGI or delta. It is the double analog
of a state of four identical proton holes ($^{96}$Pd). The $T=0$
state in the fourth column has an interesting structure with vanishing
components for $[0\,,4]$ . However, when interactions other than
$Q\cdot Q$ are used, it is no longer an eigenfunction.

In the last column, we list the other $T=2$, $v=4$ state. One sees
this on the list when one uses a seniority-conserving interaction
such as a delta interaction. However, for a general interaction, it
does not appear. This is because it gets mixed with the $T=2$, $v=2$
state. Only the state in the third column remains unscathed when we
turn on some arbitrary interaction---and only that state does not end 
up being degenerate with some other state.

There is also a unique $J=6^{+}$, $v=4$ state. With the pairing
interaction, this is degenerate with another $J=6^{+}$, $v=4$ state
and so the uniqueness gets obscured. However, with the $Q\cdot Q$
interaction, unlike the case for $J=4^{+}$, the unique $J=6^{+}$,
$v=4$ state is not degenerate with another state. Hence even with
$Q\cdot Q$ this unique state appears in the calculation.

There are other examples of confusions. The electric dipole moment
of the neutron would vanish if parity conservation holds. But at a
more important level, it vanishes if time reversal invariance holds.

\section{The first $J=1^+$, $T=0$ states in a single-$j$-shell configuration
in even--even nuclei.}

The first even--even nucleus for which there are $J=1^+$, $T=0$ states in a 
single-$j$-shell configuration is $^{48}$Cr. If we limit ourselves to single $j$, 
there are no $M1$ transitions from these states to any $J=0^+$, $T=0$ states.

\subsection{Absence of $J=1^{+}$, $T=2$ states in $j^{4}$ configurations}

In early shell model calculations by McCullen et al.~\cite{mbz64} and Ginocchio and
French~\cite{gf63}, it was noted that, in the $f_{7/2}$ shell,
certain combinations of spin and isopin did not exist. For example, there were
no $J=0^+$, $T=1$ states in $^{44}$Ti and no $J=1^{+}$ states with $T=T_{\text{min}}+1$,
where $T_{\text{min}}= |N-Z|/2$. There were also no $J=1^{+}$ states with $T=T_{\text{max}}$. 
However, those states are analogous to states of a system of identical particles, 
i.e. calcium isotopes. Explanations for some of the missing states can be shown 
by simple techniques as will be discussed later.

In the mid eighties, papers were published which counted the states in a more
systematic way. They include the works of I.~Talmi on recursion relations for 
counting the states of identical fermions~\cite{t05} and by Zhao and Arima~\cite{za05}, 
who obtained expressions for the number of $T=0$, 1 and 2 states for protons and
neutrons in a single $j$ shell. Indeed the latter authors give all the answers 
to the counting questions addressed in this paper.

\subsection{The first occurrence of $J=1^{+}$, $T=0$ states in the single $j$ 
shell---$^{48}$Cr}

There are no $J=1^{+}$, $T=0$ states for four nucleons in the single $j$ shell. 
To make things more concrete, consider $^{44}$Ti. The two $f_{7/2}$ protons can 
have angular momenta 0, 2, 4, and 6, all occurring once; likewise the two neutrons. 
The [$J_{p}$, $J_{n}$] configurations that can add up to a total $J=1$ are [2, 2], 
[4, 4] and [6, 6]. Thus, there are three $J=1^{+}$ states. The possible isospins are 
0, 1, and 2. Let us next consider $^{44}$Sc. The three neutrons can have angular 
momenta 3/2, 5/2, 7/2, 9/2, 11/2, and 15/2, all occurring only once. The states 
that add up to one are [7/2, 5/2], [7/2, 7/2], and [7/2, 9/2]. Again we have three 
states. However, since $^{44}$Sc has $|T_{z}|=1$, the isospins can only be one or 
two. Hence, there are no $J=1^{+}$, $T=0$ states in $^{44}$Ti which are of the 
$(f_{7/2})^{4}$ configuration.

We next consider $^{48}$Cr. The possible states of four protons, including
seniority labels are: $v=0$, $J=0$; $v=2$, $J=2,4,6$; $v=4$, $J=2^{*},4^{*},5,8$.
The possible $J=1^{+}$ states are [2, 2], [4, 4], [6, 6], [2$^{*}$, 2$^{*}$], 
[4$^{*}$, 4$^{*}$], [5, 5], [8,~8], [2, 2$^{*}$], [2$^{*}$, 2], [4, 4$^{*}$], 
[4$^{*}$, 4], [4, 5], [5, 4], [4$^{*}$, 5], [5, 4$^{*}$], [5, 6], [6, 5]. 
There are 17 such states with a priori possible isospins $T=0,1,2,3,4$. 
We next consider $^{48}$V, which consists of three protons and five neutrons. The 
latter can also be regarded as three neutron holes, so the possible states are 
the same for neutrons and protons. The three proton states are 3/2, 5/2, 7/2, 9/2,
11/2, and 15/2, all occurring only once. The possible $J=1^{+}$ states are [3/2, 3/2], 
[5/2, 5/2], [7/2, 7/2], [9/2, 9/2], [11/2, 11/2], [15/2, 15/2], [3/2, 5/2], [5/2, 3/2], 
[5/2, 7/2], [7/2, 5/2], [7/2, 9/2], [9/2, 7/2], [9/2, 11/2], [11/2, 9/2]. There are 14 
such states and they all must have isospins greater than zero. Hence, the number 
of $J=1^{+}$, $T=0$ states of the $(f_{7/2})^{8}$ configuration is $(17-14)= 3$.

The wave functions of these states are included in a larger compilation by
A.~Escuderos, L.~Zamick, and B.F.~Bayman ~\cite{ezb05}. It is there noted that 
because both protons and neutrons are at mid shell, the quantitiy $s=(-1)^{v}$ is 
a good quantum number, where $v=(v_p+v_n)/2$. Referring to Ref.~\cite{mbz64} for 
$J=0^{+}$, $T=0$, there are four states with $s=+1$ and two with $s=-1$. All 
$J=0^{+}$, $T=1$ states have $s=-1$, while all $T=2$ and $T=4$ states have 
$s=+1$. There are two $J=1^{+}$, $T=0$ states with $s=-1$ at energies of 7.775 
and 9.258~MeV. There is one $s=+1$ state at 9.037~MeV with a rather simple wave
function: $1/\sqrt{2} [(4^{*},5) + (5,4^{*})]$. The lowest $J=0^{+}$, $T=0$ state has 
$s= +1$.

\subsection{$M1$ selection rules}

There is a modern twist to what we are here doing. There has been an extensive 
review of $M1$ excitations, including spin-flip modes, scissors modes etc., by 
K.~Heyde, P.~Von Neumann-Cosel, and A.~Richter ~\cite{hnr10}. The mode we are
here considering has, to the best of our knowledge, not yet been studied 
experimentally. There have been studies of $M1$ $T=0 \to T=0$ transitions, e.g. 
the electro-excitation of $J=1^{+}$, $T=0$ excited states of $^{12}$C, but 
these involve more than one shell. Isospin impurities are very important 
for these transitions because the isovector $M1$ coupling constants are much 
larger than the isoscalar ones.

One simple selection rule for $M1$ transitions in this limited model space is
that $M1$ $(T=0\rightarrow T=0)$ equals zero. To see this, we note that in the 
single-$j$-shell space we can replace the $M1$ operator by $g_jJ$. The $M1$ 
matrix element for a $T=0 \to T=0$ transition is thus proportional to $(g_{j\pi}
+ g_{j\nu})$, i.e. the isoscalar sum. But if such a term is non-zero, it would 
imply that the total angular momentum operator $J$ (obtained by setting the two 
$g$'s above each equal to 1/2) could induce an $M1$ transition, which, of course,
it cannot.

Another ``midshell'' selection rule is that the quantum number $s$ has to be the 
same for the initial $J=1^{+}$, $T=0$ state and for any final state, e.g $J=1^{+}$,
$T=1$ or $J=2^{+}$, $T=1$.

Although not necessary, it is nevertheless instructive to show in more detail
why the $T=0 \to T=0$ matrix element vanishes. Consider a transition from $s=-1$
to $s=-1$. In the wave functions, there will be no amplitude of the configuration
$(J_p,J_n) =(2,2)$, but there will be of $(2,2^{*})$ and $(2^{*},2)$. The transition 
matrix element will have the form $\langle (2,2^{*})^{2} +(2^{*},2)^{2}|| M1 || 
(2,2^{*})^{1}- (2^{*},2)^{1}\rangle$. This is equal to $\langle (2,2^{*})^{2}||M1||
(2,2^{*})^{1}\rangle - \langle (2^{*},2)^{2}||M1|| (2^{*},2)^{1}\rangle$. Since
in the single $j$ shell one can replace $M1$ by $g_j J$, the matrix element
$\langle 2||M1||2\rangle$ is equal to $\langle 2^{*}|| M1 ||2^{*}\rangle$. We thus 
see that the complete matrix element vanishes.

\section{Selected Systematics of Odd--Odd Nuclear Spectra}

We consider $T=1$ states of four nucleons with three partices (holes) of one
kind and one of the other kind, e.g $^{44}$Sc (three neutrons and one proton) 
or $^{96}$Ag (three proton holes and one neutron hole). We formulated a $(2j-1)$ 
rule which will here be presented in a somewhat different way than in Ref.~\cite{ze13}.
The rule is that, for these systems, yrast states with angular momenta
$I=(2j-1)$ lie lower in energy than neighbouring states with angular momenta 
$(2j-1)-1$ or $(2j-1)+1$. We give some examples in Table~\ref{tab:2j-1}, with 
energies in MeV.

\begin{table}[h!t]
\caption{Examples of the $(2j-1)$ mentioned in the text. The energies are given
in MeV.}
\centering
\begin{tabular}{|c|c|c|c|}
\hline
 & $I$ & Exp. & Theory \\
\hline
$^{44}$Sc & 5 & 1.513 & 1.276 \\
 & 6 & 0.271 & 0.381 \\
 & 7 & 0.968 & 1.272 \\
\hline
$^{52}$Mn & 5 & 1.254 & 1.404 \\
 & 6 & 0.000 & 0.000 \\
 & 7 & 0.870 & 1.819 \\
\hline
$^{96}$Ag & 7 & ????? & 0.861 \\
 & 8 & 0.000 & 0.000 \\
 & 9 & 0.470 & 0.492 \\
\hline
$h_{11/2}$ $Q\cdot Q$ & 9 & & 1.30 \\
 & 10 & & 0.21 \\
 & 11 & & 0.85 \\
\hline
\end{tabular}
\label{tab:2j-1}
\end{table}

A possible explanation of this rule for nuclei at the end of a closed
shell is that for such nuclei the value of the rotational quantum
number $K$ is equal to $(2j-1)$. In more detail, the neutron hole has 
$k_{1}=j$ and the three proton holes have $k_{2}=j-1$, so that 
$K=k_{1}+k_{2}=(2j-1)$.

In Ref.~\cite{ze13}, it was noted that it is hard to get two-body matrix 
elements from experiment. One can get $T=1$ matrix elements from the 
spectrum of $^{98}$Cd, but the spectrum of the two-hole nucleus $^{98}$In 
is not known, so we cannot get the $T=0$ two-body matrix elements in a 
simple direct way. Sorlin and Porquet~\cite{sp08} discussed using a Pandya transformation 
to get the particle-particle spectrum from the particle-hole spectrum of $^{90}$Nb.
This is a priori a reasonable thing to try. They used as input the yrast 
spectrum of $^{90}$Nb, except for $J=1^{+}$, where they used the second
excited $1^{+}$ state. We reproduced the results in~\cite{ze13}. We find 
that for $^{96}$Cd this method gives a significantly lower excitaton energy 
for $J=16^{+}$ in $^{96}$Cd than does a realistic CCGI interaction~\cite{ccgi12}:
3.898~MeV vs 5.245~MeV. This may be due to the increasing collectivity as one
moves away from the $N=50$ $Z=50$ closed shell.

\end{document}